\documentclass[aps,prb,twocolumn,superscriptaddress,showpacs,floatfix]{revtex4}
\usepackage{graphicx}
\usepackage{amsmath}
\usepackage{amssymb}
\usepackage{url}

\newcommand{\D}{\ensuremath{{\mathrm{d}}}}
\newcommand{\E}{\ensuremath{{\mathrm{e}}}}
\newcommand{\imag}{\ensuremath{{\mathrm{i}}}}

\graphicspath{{.}{../}{./eps/}}

\begin{document}

\title{Dynamical Aspects of 2D Quantum Percolation}
\author{Gerald Schubert}
\author{Holger Fehske}
\affiliation{Institut f\"ur Physik, Ernst-Moritz-Arndt Universit\"at
  Greifswald, 17487 Greifswald, Germany}
%
\date{\today}
\begin{abstract}
The existence of a quantum percolation threshold $p_q<1$ in the 2D quantum
site-percolation problem has been a controversial issue for a long time. 
By means of a highly efficient Chebyshev expansion technique we investigate  
numerically the time evolution of particle states on finite disordered 
square lattices with system sizes not reachable up to now. After a careful 
finite-size scaling, our results for the particle's recurrence probability and
the distribution function of the local particle density give 
evidence that indeed extended states exist in the 2D percolation model for $p<1$.

\end{abstract}

\pacs{71.23.An, 71.30.+h, 05.60.Gg, 72.15.Rn}

\maketitle

\section{Introduction}

For classical percolation above a critical concentration of accessible sites, 
the so-called percolation threshold $p_c$, a spanning cluster exists, allowing 
for transport.
In the quantum case, however, the existence of the spanning cluster does not 
guarantee an extended wave function.
As for the Anderson~\cite{An58} and binary alloy~\cite{AF05} disorder models 
scattering and interference effects at the irregular boundaries of the cluster
may lead to localization of the quantum particle, i.e. absence of diffusion.
Increasing the concentration of accessible sites $p$ above $p_c$, the first 
occurrence of extended states defines a quantum percolation threshold, $p_q$.
For the Anderson model, below a critical disorder strength, a single 
mobility edge separates bands of localized and extended states.
By contrast, for the binary alloy and three-dimensional (3D) percolation model 
a sequence of ``mobility edges'' exists~\cite{SWF05}.
It is clear that $p_c \le p_q \le 1$ and the fundamental question
is, of course, whether $p_q$ equals one of these boundaries.

For the 3D case, results in the 
literature~\cite{KN90,SLG92,SWF05} agree on $p_c^{3D}<p_q^{3D}<1$, 
with estimates for $p_q^{3D}$ ranging from $0.4$ to $0.5$
for site percolation on a simple cubic lattice. 
In 2D, the situation is less clear. 
Here the physical community is evenly divided: one 
group~\cite{SG91,MDS95,BKS98,HKS02} favors 
$p_q^{2D}=1$ while another group~\cite{OC84,SC84,LZ99,DCA00,ERS01,NBR02,IN07p}
claims that $p_q^{2D}<1$.
The most striking argument against $p_q^{2D}<1$ comes from one-parameter 
scaling theory~\cite{AALR79}, according to which arbitrary small 
disorder always leads to localization in 2D.
%
%
%
Nevertheless, there are hints for $p_q^{2D}<1$. In this regard  
band center states seem to be of particular importance~\cite{ITA94,ERS98b}.
Whether $E=0$ states on a 2D bipartite depleted square lattice
are a possible ``let out clause''~\cite{GW91,Ga93} for the 
one-parameter scaling theory result is an open question.

Renewed interest in 2D quantum percolation came up in connection 
with the unusual transport properties of novel materials. For example, 
the metal-insulator transition in perovskite manganite films and the 
related colossal magnetoresistance effect seems to be inherently
percolative~\cite{ZIBGL02,BSLMDSS02}. 
Quantum percolation might be of importance for the 
experimentally observed metallic behavior of dilute weakly 
disordered Si-MOSFETs~\cite{AKS01}.  
Moreover, quite recently, a quantum percolation scenario has been 
proposed for minimal conductivity in undoped graphene~\cite{CFAA07p}. 
All this gives additional motivation for the rather fundamental study 
of 2D quantum percolation presented in this work.

\section{Model and methods}
We start from a tight-binding Hamiltonian of non-interacting spinless
fermions in Wannier representation 
\begin{equation}\label{H_bm}
  {H} =  \sum_{i=1}^{N} \epsilon_i {d}_i^{\dag} {d}_i^{} 
  - t \sum_{\langle ij \rangle}({d}_i^{\dag} {d}_j^{} + \text{H.c.})\,,
\end{equation}
with uniform hopping $t$ between nearest neighbors $\langle i j \rangle$
on a finite square lattice with $N=L^2$ sites and 
periodic boundary conditions. 
The on-site energies $\epsilon_i$ are subject to the 
bimodal distribution
\begin{equation}
  p[\epsilon_i] = p\,\delta(\epsilon_i-\epsilon_A) +
  (1-p)\, \delta(\epsilon_i-\epsilon_B)\,.
\end{equation}
The quantum site-percolation Hamiltonian then is  obtained in the limit 
$|\epsilon_B - \epsilon_A | \to\infty$. Without loss of generality
we choose $\epsilon_A=0$.
In this situation the fermions move 
on a random assembly of  $A$-lattice points. Depending on $p$, 
the largest cluster of $\bar{N}$ connected $A$-sites may span the 
entire lattice. Note that the dynamics of our finite system is governed 
by two different time scales. The time scale of a hopping process is 
given by the inverse of the hopping matrix element, $\tau_0 = 1/t$. 
To account for the system size, we define a characteristic time 
$\bar{T}_0=\bar{N}\tau_0$, which, in principle, is necessary to visit
each site of the spanning cluster once.

In order to address the problem of localization in quantum 
percolation, we determine, for $p>p_c=0.593$, the recurrence 
probability of a particle to a given site, $P_R(\tau)$, which 
for $\tau\to\infty$ may serve as a criterion 
for Anderson localization~\cite{KM93,Ja98}. 
While for extended states on the spanning cluster 
$P_R(\tau\to\infty)=1/\bar{N}$ scales to zero in the thermodynamic limit, 
for localized states $P_R(\tau)$ remains finite as $\bar{N}\to\infty$.

Conceptionally, in a first step, we track the time evolution of an 
initially localized state (which, of course, in general is not an eigenstate) 
by calculating the modulus square of the wave function 
at the particle's starting position as a function of time $\tau$. 
To this end we expand the time evolution operator 
$U(\tau) = \E^{-\imag H\tau}$ in Chebyshev polynomials~\cite{TK84,CG99,WF08},
%
%
where the Hamiltonian has to be rescaled to 
the definition interval of the Chebyshev polynomials $T_k$, $[-1,1]$. 
With $H=a\tilde{H}+b$ we obtain

\begin{equation}
  U(\tau) =  \E^{-\imag b \tau} 
  \left[ c_0(a\tau) + 2\sum\limits_{k=1}^{M} c_k(a\tau) T_k(\tilde{H}) \right]\;.
\end{equation}

The expansion coefficients $c_k$ are given by 
\begin{equation}
  c_k(a\tau) = \int\limits_{-1}^1 \dfrac{T_k(x)e^{-\imag a\tau x}}{\pi \sqrt{1-x^2}}\D x =
  (-\imag)^k J_k(a\tau)\;,
\end{equation}
where $J_k$ denotes the Bessel function of the first kind. 
To calculate the evolution of a state $|\psi(\tau)\rangle$ from one 
time grid point to the next,
$|\psi(\tau+\Delta\tau)\rangle = U(\Delta\tau)|\psi(\tau)\rangle$, we
have to accumulate the $c_k$-weighted contributions of the different orders 
$|v_k\rangle = T_k(\tilde{H})|\psi(\tau)\rangle$.
As the coefficients $c_k(a\Delta\tau)$ depend on the time step but not
on time explicitly, we need to calculate them only once.
The contributions $|v_k\rangle$ can then be calculated iteratively using
the recurrence relation of the Chebyshev polynomials
\begin{equation}
    |v_{k+1}\rangle = 2\tilde{H} |v_k\rangle - |v_{k-1}\rangle\;,
\end{equation}
where $|v_1\rangle = \tilde{H} |v_0\rangle$ and $|v_0\rangle = |\psi(\tau)\rangle$.
Due to the fast asymptotic decay of the Bessel functions
\begin{equation}
  J_k(a\tau)
  \sim \frac{1}{\sqrt{2\pi k}} \left( \frac{\E a\tau}{2k} \right)^k 
  \quad \text{for}\quad k\to \infty\;,
\end{equation}
the coefficients $c_k$ vanish rapidly above a certain expansion order.
As a result from the infinite series only a finite number of $M$ terms  needs
to be taken into account. Clearly $M$ depends on the time step used. 
Truncating the series for large $a\tau>10^3$ at $M\sim 1.5 a\tau$, 
the $|c_k|<10^{-16}$ for all discarded terms.  
For even larger values of $a\tau$, the necessary value of $M/(a\tau)$
is even reduced and approaches $M\sim a\tau$ for $\tau\to\infty$.
Thus, as compared to the standard Crank-Nicolson algorithm, 
the Chebyshev expansion permits the use of a considerably larger 
time step to achieve the same accuracy, i.e. allows for a very efficient
calculation of the time evolution of a given state.

In a second step, we employ the so-called local distribution 
approach~\cite{AF06,AF08}.
%
%
In the theoretical investigation of disordered systems it turns out
that the distribution of random quantities is central. 
While all characteristics of a certain material are determined by the
distribution $p[\epsilon_i]$, each actual sample 
constitutes only one particular realization $\{\epsilon_i\}$.
At each randomly chosen site $i$ of such a particular sample, we observe 
different local environments. That is disorder breaks 
translational invariance and   
we have to focus on site-dependent quantities like 
the local density of states (LDOS)~\cite{SWWF05} at site $i$
\begin{equation} \label{LDOS}
  \rho_i(E) = \sum\limits_{m=1}^{\bar{N}}
  | \langle i | m \rangle |^2\, \delta(E-E_m)\;.
\end{equation} 
Given an energy $E$, the LDOS can be efficiently calculated 
by means of the kernel polynomial method~\cite{WWAF06}. 
Considering the LDOS, a well established criterion 
for localization is the following. 
Probing different sites in the sample and recording the values of 
$\rho_i$ gives the probability distribution $f[\rho_i]$. 
To alleviate the problem of statistical noise one usually considers 
instead of $f[\rho_i]$ the distribution function 
\begin{equation}
  F[\rho_i]=\int_0^{\rho_i} f[\rho_i']\,\D \rho_i'\;.
\end{equation}
In the thermodynamic limit both $f[\rho_i]$ and $F[\rho_i]$ are 
independent of the actual realization $\{\epsilon_i\}$ 
(i.e. self-averaging), but solely depend on $p[\epsilon_i]$.
Therefore $F[\rho_i]$ characterizes $H(p[\epsilon_i])$, not only 
$H(\{\epsilon_i\})$. That means, at the level of distributions, we have
restored translational invariance.  
For an extended state the amplitude of the wave function is 
more or less uniform and $f[\rho_i]$ is sharply peaked and 
symmetric around its mean value $\langle \rho_i \rangle$. 
For localized states $\rho_i$ strongly fluctuates throughout the lattice, 
giving rise to a very asymmetric $f[\rho_i]$ with a long tail and 
$\langle \rho_i\rangle \to 0$. Consequently the distribution function 
$F[\rho_i]$ steeply rises for extended states whereas for 
localized states the increase extends over several orders of magnitude.
These arguments also hold for the (time-dependent) 
particle density at site $i$ 
\begin{equation}
  n_i(\tau) = \Big| \sum\limits_{m=1}^{\bar{N}}  
    \E^{-\imag E_m t} \langle m| \psi(0)\rangle \langle i| m \rangle \Big|^2\;,
\end{equation}
if we take additional care of some aspects. 
Evolving an arbitrary initial state $|\psi(0)\rangle$ in time, we only
have access to $n_i(\tau)$, containing contributions of the whole spectrum. 
Calculating the time evolution of an initial state
 we have access to $n_i$ on the whole cluster,
which gives a much better statistics. On the other hand, 
the chosen initial conditions introduce an additional dependency.
Especially for localized states the local environment of the starting
position will influence $F[n_i]$. So for $n_i$,  in addition to taking 
the thermodynamic limit, we have to examine different initial conditions.
While the LDOS approach allows for an energy resolved examination 
of the localization problem, $n_i$ gives no information about $p_q(E)$ 
because it contains contributions from all $E$.
But, starting from $p_c$ and increasing $p$, we detect at a certain 
$p=p_q$ the first occurrence of extended states somewhere in the spectrum.

\section{Results}

\begin{figure}
  \centering 
  \includegraphics[width=\linewidth,clip]{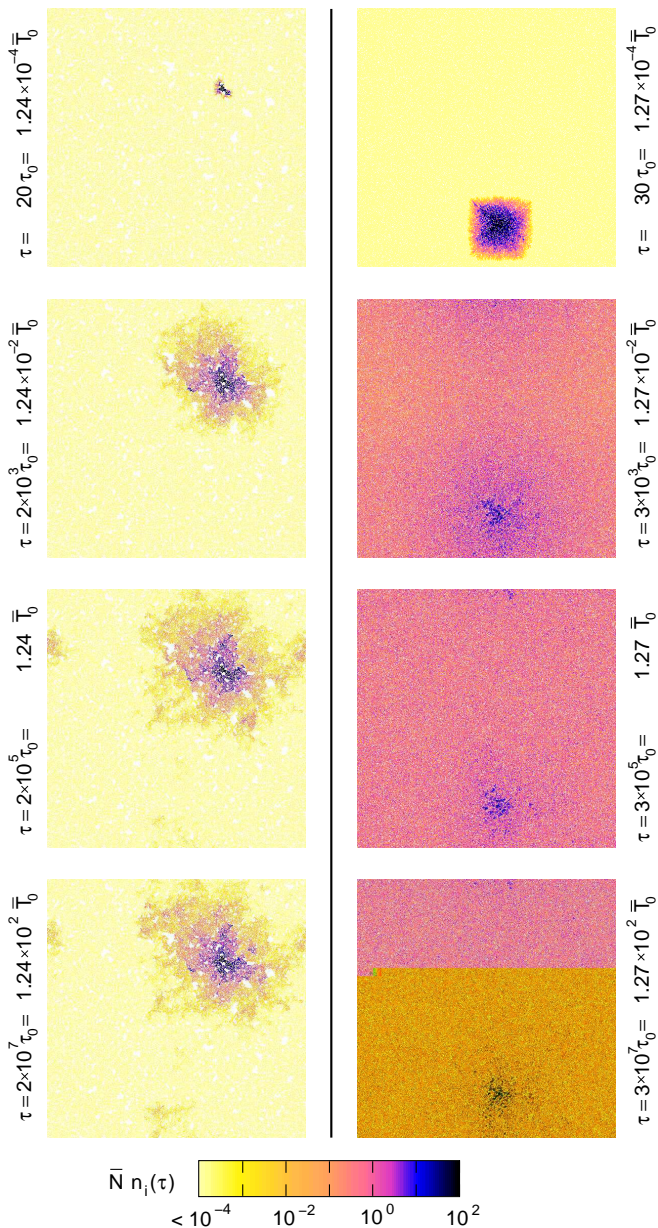}
  \caption{(Color online) Time evolution of the normalized local particle 
    density $\bar{N}n_i(\tau)$ for an initially localized state 
    on the spanning cluster
    of a $N=512^2$ square lattice for $p=0.65$ (left) and $p=0.90$ (right).
    Due to normalization, for an extended
    state evenly spread over all sites of the spanning cluster this
    quantity is equal to unity.}
  \label{perc_2d_waveampl}
\end{figure}

\begin{figure}
  \centering 
  \includegraphics[width=\linewidth,clip]{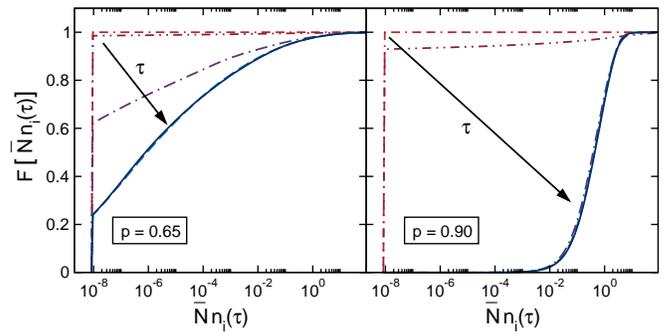}
  \caption{(Color online) Time evolution of the distribution 
    function of the local particle density, $F[\bar{N}n_i(\tau)]$. The
    curves correspond to the data in Fig.~\ref{perc_2d_waveampl}, 
    supplemented by the distribution functions at $\tau=0$ 
    (dot-double-dashed curve).
  }
  \label{DistEvolve}
\end{figure} 

Let us now discuss the outcome of our numerical investigations. 
Figure~\ref{perc_2d_waveampl} compares the time evolution 
of an initially localized state on the spanning cluster for 
low (left panels) and high (right panels) concentration of $A$
sites, for which qualitatively different behaviors emerge.  
For small $p=0.65>p_c$, the initial state spreads within a short 
time over a finite region of the spanning cluster. But 
also for very long times, $\tau>100\bar{T}_0$, 
this extension does not change significantly anymore.
For larger $p$, the spreading is even faster.
In contrast to the previous case, for $p=0.90$ the state is not confined
to some finite region, but $n_i$ is transfered to the whole cluster. 
Since the initial state is a superposition of eigenstates from the 
whole spectrum, it also contains contributions from 
localized states, for instance near the band edges. 
Those are reflected in the darker spots in the 
vicinity of the initial position, which 
persist there, even for very long times. 
This signature of localized states we already know 
from the case of small $p$. The essential new feature is 
that for large $p$ some eigenstates exist, which are not localized, 
i.e. spread over the whole cluster.

To go beyond this simple visualization and account for  
a quantitative description, in Fig.~\ref{DistEvolve} we show 
the corresponding local particle density distribution functions
$F[n_i]$. The different behavior of localized and extended states 
becomes obvious as the time increases. 
Since these results were obtained for a particular sample and 
a certain starting position it is necessary to ask how strong
these findings depend on the initial conditions, and if they are 
representative for the underlying $p[\epsilon_i]$.

To answer this question we examine systematically the influence of the initial
conditions on the behavior of the state.
In this respect the geometry and energy of the initial state, 
the starting position on the cluster as well as the particular 
cluster realization $\{\epsilon_i\}$ should be of importance. 
A fingerprint in which all these aspects come into play 
is the LDOS of the initial state.

\begin{figure}
  \centering 
  \includegraphics[width=\linewidth,clip]{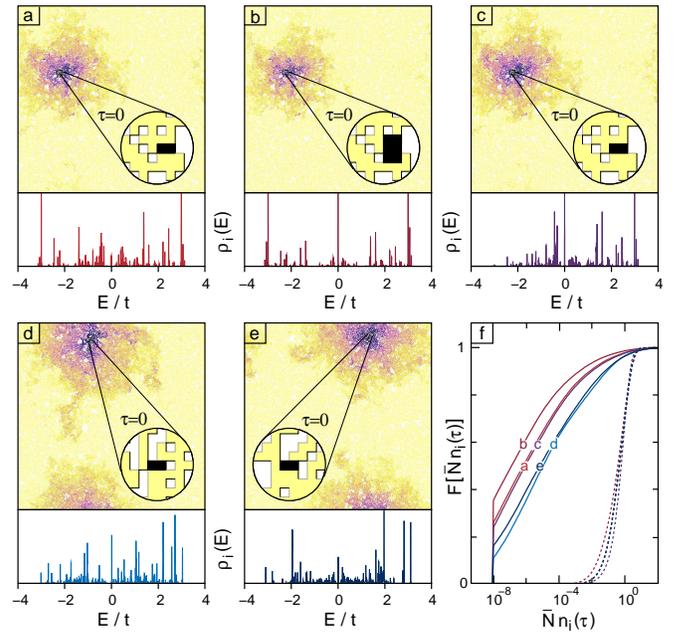}
  \caption{(Color online) Influence of the initial conditions on the evolution
    of the wave function on a $N=512^2$ system at $p=0.65$. 
    The colormap is the same as the one in Fig.\ref{perc_2d_waveampl}.
    Panels a-d show the same cluster realization, 
    with identical starting positions for a-c.
    All states have $E=0.5t$, except for the case shown in panel c, 
    where $E=t$. The insets give a
    magnification of $n_i(0)$. The corresponding LDOS
    of the initial states are given below each panel. 
    In panel f the solid lines show the distribution function $F[n_i(\tau)]$
    for $\tau=15.4\bar{T}_0$.
    For comparison the dashed lines indicate $F[n_i(\tau)]$ for $p=0.90$ 
    subject to the same differences in the initial conditions.}
  \label{Dependency_IC}
\end{figure}

In Fig.~\ref{Dependency_IC} we examine the influence of these issues 
one by one. Fixing a particular cluster realization and starting position 
of the wave function, we may construct several initial states with 
the same total energy. 
This can be done by choosing a different number of sites 
which initially have non-vanishing amplitudes.
For a state with finite amplitude only on two neighboring sites, $a$ and 
$\sqrt{1-a^2}$, respectively, we get $E = 2ta\sqrt{1-a^2}$ and we may 
continuously tune $E\in[-t,t]$ by choice of $a$ 
(cf. Fig.~\ref{Dependency_IC} a and c).  
These energies may also be constructed for more complicated initial states, 
as e.g. for the one shown in Fig.~\ref{Dependency_IC} b, 
with non-vanishing amplitudes on six adjacent sites.
The more complicated these structures get and the more sites are involved, the
larger energies of the initial state are possible, e.g. 
for the configuration of Fig.~\ref{Dependency_IC} b up to $(1+\sqrt{2})t$.
As stressed above, these energies are not eigenenergies
of $H$, i.e. the initial state is a superposition of eigenstates from the
whole spectrum. A quantitative characterization of the amount 
of different contributions 
can be obtained by the LDOS of the initial state.
Changing the starting position (Fig.~\ref{Dependency_IC} d) or the cluster 
realization (Fig.~\ref{Dependency_IC} e) has a similar effect. 
Essentially, we see the same behavior as for the changes in 
Figs.~\ref{Dependency_IC} a-c.
Most notably, although the LDOS at $\tau=0$ and the 
evolution of the states differs in detail,
on a coarse grained scale they behave similarly. 
This is also corroborated by the behavior of the 
distribution function $F[n_i(\tau)]$ for sufficient long times $\tau$.
Despite minor differences between the five curves, they agree on showing the 
characteristics of a localized state 
(Fig.~\ref{Dependency_IC} f, solid lines). 
The more a state is localized, i.e. the lower $p$, the more the local structure
of the spanning cluster influences its evolution and thus $F[n_i(\tau)]$.
At large $p$, the dependency of an extended state on the
above criteria is much weaker, as in this case the local environment of the 
starting position is less important (Fig.~\ref{Dependency_IC} f, 
dashed lines).  Overall, the characteristics of the time 
evolution is mainly determined by the cluster structure, 
the initial state has only minor influence. 
This holds as the random nature of the cluster guarantees 
a similar structure above a certain scale.
Finally one may ask if the chosen boundary conditions or the linear extension
(odd/even) of the lattice influences the dynamics.
The latter distinction has been shown to be 
important for a bond percolation model because of the bipartiteness
of the underlying lattice~\cite{SWGE82}. 
In our case, however, we did not see an influence on the results.

%
%
\begin{figure}
  \centering 
  \includegraphics[width=\linewidth,clip]{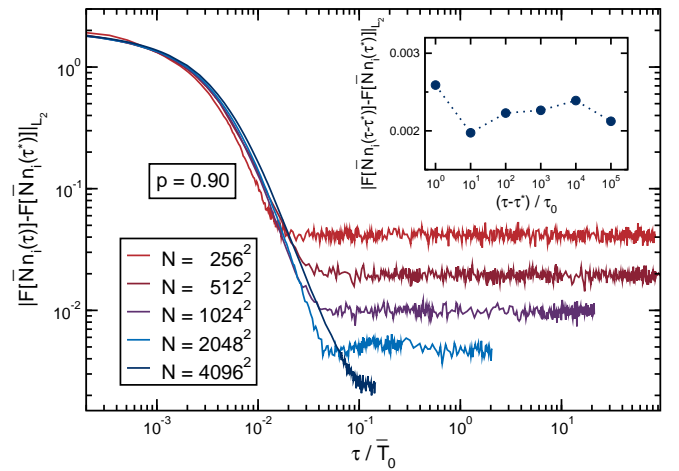}
    \caption{(Color online) $L_2$ norm of the difference between the 
      distribution function $F[n_i(\tau)]$ at time $\tau$ and the
      quasi-stationary distribution $F[n_i(\tau^*)]$ for different system
      sizes. The inset shows the dependency of this quantity on the time difference
      for the $N=4096^2$ system in the quasi-stationary regime.
      Note the different axis scales.
      }
\label{DistDiff}
\end{figure} 
In view of a finite-size scaling of the numerical data we first have to 
ensure that the obtained distribution function has already become 
quasi-stationary. 
To check this, we calculate the $L_2$ norm of the difference between two
distribution functions at different times.  
Due to fluctuations, we cannot expect this quantity to vanish completely
even for large times.
For $\tau\gtrsim \tau^* \approx 0.1\bar{T}_0$, however, we have 
reached the quasi-stationary regime for all considered 
system sizes (cf. Fig.~\ref{DistDiff}). 
At this time, the wave function has reached its maximum extension and the
further development is governed by amplitude fluctuations from site to 
site. 
In this regime, the difference between the distribution 
functions does not depend on time anymore (inset of Fig.~\ref{DistDiff}),
which is a clear indication of random fluctuations without additional drift motion.
This allows for a determination of the quasi-stationary distribution 
function together with an error estimation and enables us to 
compare different system sizes. 
In view of computation time, the scaling $\tau^* \sim \bar{T}_0 \sim \bar{N}$
together with the linear dependency on the dimension of the Hilbert space 
for the Chebyshev expansion leaves us with a 
scaling $\sim (pN)^2 \sim p^2 L^4$. 

\begin{figure}
  \centering 
  \includegraphics[width=\linewidth,clip]{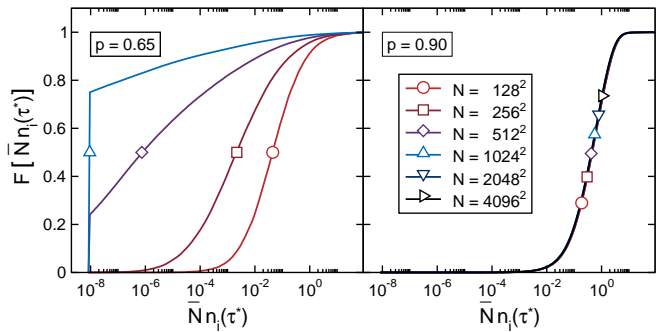}
    \caption{(Color online) Finite-size scaling of $F[\bar{N}n_i(\tau^*)]$ in 
      the quasi-stationary regime.} 
\label{FSS_Dist}
\end{figure}

As soon as we have quasi-stationary distribution functions, their dependency
on the system size may be exploited for a quantitative distinction between 
localized and extended states (Fig.~\ref{FSS_Dist}). 
For the two different occupation probabilities, we observe completely
different behavior: while for $p=0.65$ the distribution 
function shifts towards smaller values on increasing the 
system size, for $p=0.90$ it is not affected by the change of 
system size at all. The latter is clearly the behavior one 
would expect for an extended state.

\section{Conclusion}
To summarize, we applied a highly efficient Chebyshev expansion technique to calculate the 
dynamics of particle states for the 2D quantum site-percolation model.
The local particle density contains contributions of eigenstates from the 
whole energy spectrum and is therefore an ideal tool to globally investigate the
localization properties of the system.
Examining the corresponding distribution function allows for a
distinction between localized and extended states.
Above a certain concentration of accessible sites the shape of the distribution
function becomes independent of the system size.
This gives evidence that there are (some) extended states in the spectrum.
Having access to much larger systems than any others previously in the literature,
we conclude from our data, supplemented by a careful finite-size scaling, 
that $p_q^{2D}<1$.

\subsection* {Acknowledgments}
We thank J. Kantelhardt for helpful discussion. The numerical 
calculations have been performed on the HLRB at LRZ 
Munich and the TeraFlop compute cluster at the Institute of 
Physics, Greifswald University.



\end{document}